\documentclass[12pt]{article}
\pdfoutput=1
\usepackage{graphicx,subfigure,epstopdf}
\usepackage{amssymb,amsmath,cite,dsfont}
\usepackage{latex8}              
\begin{document}

\title{Consequences of Many-cell Correlations in Treating Clocked Quantum-dot Cellular Automata Circuits}

\author{Marco Taucer, Robert A. Wolkow\\
Department of Physics\\ University of Alberta \\ Edmonton, AB, CANADA\\ \{taucer, wolkow\}@ualberta.ca\\
\and
Faizal Karim, Konrad Walus\\
Department of Electrical\\and Computer Engineering\\
The University of British Columbia\\ Vancouver, BC, CANADA\\
\{faizalk, konradw\}@ece.ubc.ca\\
}
\maketitle

\begin{abstract}
Quantum-dot Cellular Automata (QCA) provides a basis for classical computation without transistors. Many simulations of QCA rely upon the so-called Intercellular Hartree Approximation (ICHA), which neglects the possibility of entanglement between cells. Here, we present computational results that treat small groups of QCA cells with a Hamiltonian analogous to a quantum mechanical Ising-like spin chain in a transverse field, including the effects of intercellular entanglement. When energy relaxation is included in the model, we find that intercellular entanglement changes the qualitative behaviour of the system, and new features appear. In clocked QCA, isolated groups of active cells experience oscillations in their polarization states as information propagates. Additionally, energy relaxation tends to bring groups of cells to an unpolarized ground state. This contrasts with the results of previous simulations which employed the ICHA. The ICHA is a valid approximation in the limit of very low tunneling rates, which can be realized in lithographically defined quantum-dots. However, in molecular and atomic implementations of QCA, entanglement will play a greater role. The degree to which entanglement poses a problem for memory and clocking depends upon the interaction of the system with its environment, as well as the system's internal dynamics.
\end{abstract}

\section{Introduction}
Quantum-dot cellular automata (QCA) provides a basis for binary computation that is fundamentally different from today's transistor-based technology~\cite{Lent:1993, Tougaw:1994}. With QCA, information is stored and transferred by cells of quantum dots, such as those depicted in Figure~\ref{fig:twostate}. A cell consists of four quantum dots arranged at the corners of a square. Each cell contains two mobile electrons, which can tunnel between dots. The electrons experience mutual repulsion and at sufficiently low temperature will take either of the two possible antipodal configurations (or a superposition thereof). Because of the symmetry of the cell, neither configuration is preferred in the absence of a perturbation, and the cell, with its two electrons, can be thought of as a bistable switch: a small push toward one or the other diagonal will cause the electrons to ``switch'' almost completely. The interaction between cells tends to align the electronic configuration of one cell with that of its neighbours. A line of cells thus acts as a binary wire. Different geometrical arrangements of cells correspond to logic gates, and together enable the design of a universal computer capable of very low power operation~\cite{Timler:2002, Liu:2006}. Computing speed would be enormously improved by the incorporation of clocking\cite{Blair:2003-2, Amlani:2000}, whereby clock zones are used to sequentially send ``bit packets'' through a QCA array to allow for pipelining. A clock zone consists of a group of cells whose parameters can be independently tuned to either allow or disallow the group of cells to become polarized (\textit{i.e.}, to take on one of the antipodal configurations). A bit packet consists of a group of interacting cells in ``active" clock zones. As envisioned theoretically, many bit packets could be processed simultaneously in a single QCA circuit by modulating clock zone parameters appropriately.

QCA cells were demonstrated experimentally in metal-island quantum dots as early as 1997~\cite{Orlov:1997}. These metal-island QCA cells have been used as prototypical devices to demonstrate transmission of information, logic gates~\cite{Amlani:1999, Snider:1999}, and clocking~\cite{Orlov:2001, Orlov:2003}.  In this implementation, clocking is realized by modulating the null-dot potential between dots. However, the relatively large size of the quantum dots means that energy levels are closely spaced, so metal-island devices must be kept at temperatures below $\sim 5K$~\cite{Snider:2003, Snider:2004} in order for quantum effects to be observable. As the dimensions of a QCA cell are reduced, the operating temperature increases, and at the molecular scale room temperature operation becomes possible. Further advantages of miniaturization include fast switching times and increased device density. For these reasons, molecular scale QCA has held great promise. Suitable candidates have been synthesized~\cite{Jiao:2003, Lu:2007}, and a QCA cell made of atomic quantum dots on silicon has also been realized\cite{Wolkow:2009}. For QCA cells of this size it is probably neither feasible nor desirable to have addressable control over the parameters of individual cells. Instead, large groups of cells could be addressed by external fields\cite{Blair:2003-2, Hennessy:2001, Faizal:2010}. The variation in these external fields can produce clock zones.

A great deal of theoretical and modelling work has been done on the topic of QCA\cite{Lent:1993-3, Tougaw:1994,Tougaw:1996, Toth:2001-2, Faizal:2008, My:IEEEProceedings:2006}. This research has aimed to capture the qualitative and quantitative characteristics of QCA cells and of arrays of cells. Because of the difficulties inherent in solving the complete quantum mechanical problem, a number of simplifying assumptions are typically made. These include a reduction of the Hilbert space to two states per cell~\cite{Tougaw:1996}, treatment of intercell interactions via a mean-field approach~\cite{Tougaw:1994, Lent:1993-3}, and finally an assumption of exponential energy relaxation~\cite{Toth:2001-2, Timler:2002}. Although much insight has already been gained on the nature and importance of quantum mechanical calculations that go beyond these approximations, including full and partial quantum correlations~\cite{Tougaw:1996, Toth:2001-2}, some important features of clocked QCA systems remain unreported. We have found that the inclusion of intercell correlations can significantly change the steady state of the system not only quantitatively, but qualitatively, even in the case of very simple systems, such as an unbiased line of QCA cells. Full quantum mechanical simulations, with the approximation of exponential relaxation to a thermal steady-state, predict an exponential loss of information even as bit packets propagate, as well as coherent oscillations whose period strongly depends on the size of the bit packet. This contrasts with mean-field simulations, which show indefinite propagation of information. The results presented here will have implications for molecular-scale QCA device design, and will highlight the need for implementation-specific theoretical treatments of the interaction of a QCA system with its environment.

In Section \ref{sec:2} we review the basic theory of QCA as well as some of the most common approximations used in QCA simulations. In Section \ref{sec:qm} we consider the full Hamiltonian for a QCA line, and the characteristics of its solutions. In Section \ref{sec:depol} we present the results of fully quantum mechanical simulations. Section \ref{sec:disc} discusses the main results of the paper, their scope, and their implications for QCA design, and finally we offer a conclusion in Section \ref{sec:conc}. 

\section{Simulation of QCA Systems} \label{sec:2}
The dynamic behaviour of QCA was first explored in~\cite{Tougaw:1996}, where Tougaw~\textit{et~al.} examined the time evolution of QCA cells by considering a basis set consisting of all sixteen possible states of a four-dot QCA cell populated by two electrons of opposite spin. In this sixteen-state approach, the authors construct a Hubbard-type Hamiltonian, given as~\cite{Tougaw:1996} 

\begin{eqnarray}
\hat{H} &=& \sum_{i, \sigma,m}E_{0}\hat{n}_{i,\sigma}(m) - \sum_{i>j,m,\sigma} t_{i,j} \left[ \hat{a}^{\dagger}_{i,\sigma}(m)\hat{a}_{j,\sigma}(m) + \hat{a}^{\dagger}_{j,\sigma}(m)\hat{a}_{i,\sigma}(m)\right] 
+ \sum_{i,m}E_Q\hat{n}_{i,\uparrow}(m)\hat{n}_{i,\downarrow}(m) \nonumber \\ &+& \sum_{i>j,\sigma,\sigma^{'},m}V_Q\frac{\hat{n}_{i,\sigma}(m)\hat{n}_{j,\sigma^{'}}(m)}{|r_i(m) - r_j(m)|}
+ \sum_{i,j,\sigma,\sigma^{'},k>m}V_Q\frac{\hat{n}_{i,\sigma}(m)\hat{n}_{j,\sigma^{'}}(k)}{|r_i(m) - r_j(k)|},
\label{eqn:hubbard}
\end{eqnarray}

where the operator $\hat{a}_{i,\sigma}(m)$ ($\hat{a}^{\dagger}_{i,\sigma}(m)$) annihilates (creates) an electron on the $i^{th}$ site of cell $m$ with spin $\sigma$, the operator $\hat{n}_{i,\sigma}(m) \equiv \hat{a}_{i,\sigma}^{\dagger}(m) \hat{a}_{i,\sigma}(m)$ is the number operator for an electron on the $i^{th}$ site of cell $m$ with spin $\sigma$, and $V_Q = q_e^2/(4 \pi \epsilon)$ is a constant where $q_e$ is the charge of the electron and $\epsilon$ the electrical permittivity of the medium. The first term in Eq.~\ref{eqn:hubbard} represents the on-site energy of a dot. The second term describes the electron tunnelling, where $t_{i,j}$ is a hopping constant (with units of energy) between neighbouring sites $i$ and $j$, determined from the structure of the potential barriers between the dots in the cell. The third term in Eq.~\ref{eqn:hubbard} accounts for the energetic cost, $E_Q$, of putting two electrons of opposite spin at the same site, and the final two terms are related to the Coulombic interactions between electrons in the same cell and in neighbouring cells, respectively. The polarization of each cell can then be found by evaluating

\begin{equation}
P_m = \frac{(\rho_{1}^m + \rho_{3}^m) - (\rho_{2}^m - \rho_{4}^m)}{\rho_{1}^m + \rho_{2}^m + \rho_{3}^m + \rho_{4}^m},
\end{equation}
where $\rho_{i}^m$ is the expectation value of the number operator on the $i^{th}$ site of cell $m$; \textit{i.e.}, $\rho_{i}^m~=~\langle \hat{n}_i(m) \rangle$. Sites within a cell are labeled clockwise starting from the top right, as shown in Figure \ref{fig:twostate}. While this Hamiltonian considers the complete many-body configuration space, including correlation effects within and between cells, it becomes computationally intractable when used to model large systems. For example, a three cell system requires $16^3 = 4096$ basis vectors. The exponential growth of the basis set makes it computationally prohibitive to model any circuit larger than just a few cells. We will now discuss three ubiquitous approximations used for studying QCA circuits and systems.\\

\subsection{Two-State Approximation} \label{sec:two_state}
\begin{figure}[ht]
\centering \scalebox{0.35}{
\includegraphics{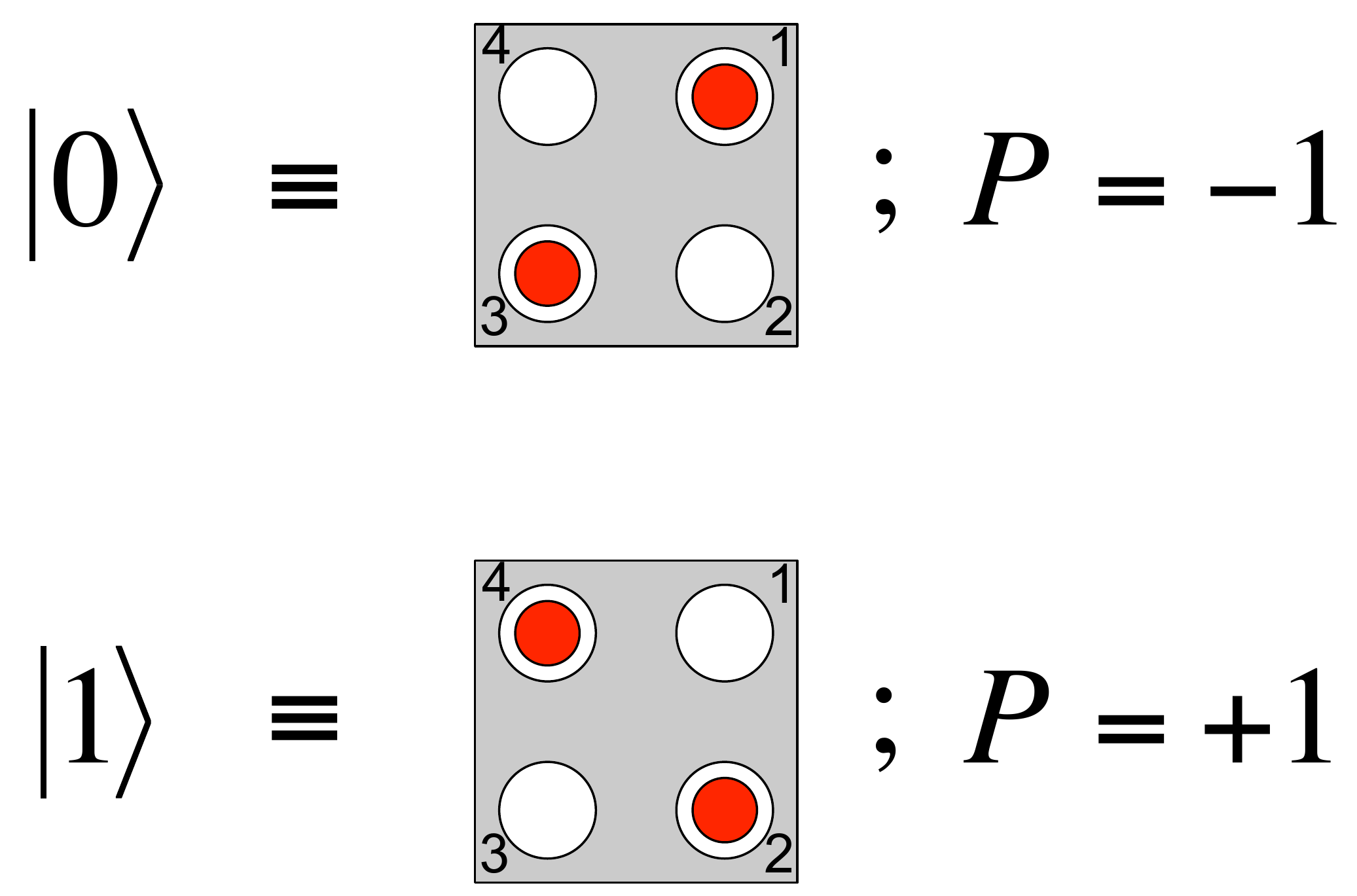}}
\caption{Schematic representation of the two diagonal states which form the basis in the two-state approximation.}
\label{fig:twostate}
\end{figure}

As shown in~\cite{Tougaw:1996}, the ground state of a single cell within the full sixteen-dimensional Hilbert space remains almost completely contained within the two-dimensional subspace of the polarized basis vectors. For sufficiently low temperatures, we can therefore expect the state of a cell or a line to be well described by the two-state approximation. We refer to this reduced basis as the polarization basis, and denote the two states as $| 0 \rangle$ and $| 1 \rangle$, as shown in Figure~\ref{fig:twostate}. The Hamiltonian in the polarization basis for a system of $N$ interacting QCA cells, under the influence of driver cells, is then described by a $2^N\times2^N$ Ising-like Hamiltonian:

\begin{eqnarray}
\hat{H} = - \sum_{i=1}^{N} \gamma_i \hat{\sigma}_x(i) - \frac{1}{2}\sum_{i<j}^{N}E_k^{i,j}{\hat\sigma}_z(i)\hat{\sigma}_z(j) + \frac{1}{2}\sum_{D}\sum_{i=1}^{N}E_k^{i,D}P_D {\hat\sigma}_z(i),
\label{eqn:ising}
\end{eqnarray}
where $\gamma_i$ is an effective tunneling energy, related to the hopping energy, $t_{i,j}$, in equation \ref{eqn:hubbard}. $E_k^{i,j}$ is the so-called kink energy between cells $i$ and $j$, and accounts for the energetic cost of two cells having opposite polarization. $P_D$ labels the polarization of the driver cell labelled $D$, and $E_k^{i,D}$ is the kink energy between cell $i$ and driver $D$. The driver cells provide a mechanism for input into the QCA circuit and have a polarization that can range from $-1$ to $+1$. The Pauli operators for the $i^{th}$ cell, $\hat{\sigma}_a(i)$; $a=x,y,z$, represent the tensor product of $N$ $2 \times 2$ identity operators, with the $i^\text{th}$ identity operator replaced by the one of the Pauli matrices,

\begin{equation} \nonumber
\hat{\sigma}_x= \left(\begin{array}{cc} 
0 & 1\\ 
1 & 0
\end{array}\right), \quad
\hat{\sigma}_y= \left(\begin{array}{cc} 
0 & i\\ 
-i & 0
\end{array}\right), \quad
\hat{\sigma}_z= \left(\begin{array}{cc} 
-1 & 0\\ 
0 & 1
\end{array}\right).
\label{eqn:pauli}
\end{equation}
\newline
For example, $\hat{\sigma}_y(2)\equiv\mathds{1} \otimes \hat{\sigma}_y \otimes \mathds{1} \otimes \ldots \otimes \mathds{1}$. The polarization of a cell, $i$, can now be defined as $P_i = -\langle \hat{\sigma}_z(i) \rangle$. The first term in Eq.~\ref{eqn:ising} accounts for the kinetic energy of electrons, and tends to bring the cells to a superposition of polarization states. The second and third terms account for the energy cost of having a cell misaligned with its neighbour, or with a driver cell respectively. The Hamiltonian in Eq.~\ref{eqn:ising} is very similar to the quantum Ising model with a transverse field, which has been extensively studied\cite{Lieb:1961, Pfeuty:1969}.

In Section \ref{sec:qm} we will use the two-state approximation to study systems of interacting QCA cells. This procedure, like the full sixteen-state one, naturally includes the effects of inter-cell entanglement, or quantum correlations. The price we pay for proper inclusion of quantum correlations is an exponential growth in the basis set with the number of interacting cells, $N$. Even with the two-state approximation, this limits its application to systems containing only a small number of interacting cells. Further approximations are used to solve the problem of scaling.

\subsection{Intercellular Hartree Approximation} \label{sec:ICHA}
One way to eliminate the problem of exponential scaling is to ignore inter-cellular entanglement effects altogether and solve the Schr\"{o}dinger equation for each individual cell separately. This method is known as the intercellular Hartree approximation (ICHA)~\cite{Tougaw:1994,Lent:1993-3}. In this Hartree-type treatment, cells are coupled to one another via expectation values (polarizations) rather than operators. The Hamiltonian (in the polarization basis) for a single cell $i$, is then simply

\begin{equation}
\hat{H}_i=-\gamma_i \hat{\sigma}_x + \frac{1}{2}\sum_{j}E_k^{i,j}P_j \hat{\sigma}_z,
\label{eqn:singlecell}
\end{equation}
where $P_j$ is the polarization of cell $j$. The polarization of cell $i$ is found by evaluating, $P_i = -\langle \sigma_z \rangle$. Because the solutions of one cell's Hamiltonian define parameters that enter its neighbours' Hamiltonians, the system of Schr\"{o}dinger equations must be solved iteratively to obtain self-consistency. In calculating the state at a particular time, the initial guess is typically taken to be the state of the system at the previous time step.

\begin{figure}[ht]
\centering
\scalebox{0.3}{
\includegraphics{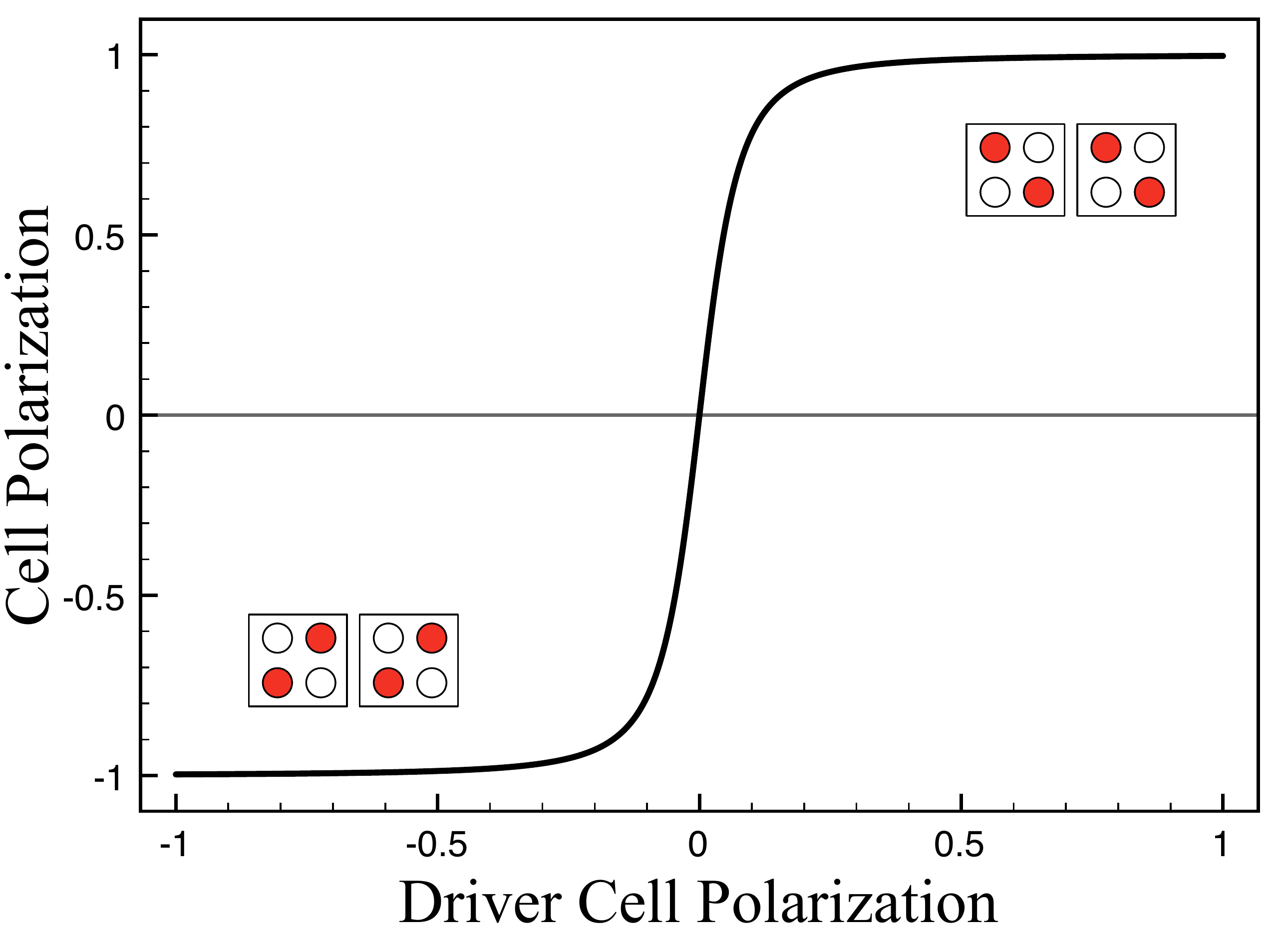}}
\caption{Nonlinear cell-to-cell response function for $\gamma \ll E_k$ and $k_B T \ll E_k$ (optimal conditions for QCA). The output cell is almost completely polarized for even a small driver polarization.}
\label{fig:cell2cell}
\end{figure}

The primary benefit of this approximation is that it only requires the diagonalization of $N~2\times2$ Hamiltonians, which scales linearly with the number of cells in the system. Furthermore, if a circuit evolves adiabatically and cells are assumed to remain at the ground state, then this problem simplifies further by recognizing that the polarization of any cell, $i$, can be evaluated analytically using~\cite{Lent:1993-2},

\begin{equation}
P_i = \frac{\frac{1}{2\gamma}\sum\limits_j E_k^{i,j}P_j}{\sqrt{1+\left(\frac{1}{2\gamma}\sum\limits_j E_k^{i,j}P_j\right)^2}}.
\label{eqn:polarization}
\end{equation}
Eq.~\ref{eqn:polarization} produces the well-known nonlinear cell-to-cell response function shown in Figure~\ref{fig:cell2cell}. While the ICHA is generally capable of arriving at the correct ground state of an array of QCA cells in a single clocking zone with a fixed driver, it can falsely predict a latching mechanism within a group of cells that allows them to retain (and even obtain) a polarization in the absence of a perturbing cell. This goes beyond previously noted inaccuracies of the ICHA, such as in dynamics and finite temperature behaviour\cite{Tougaw:1996}, where the many-cell excited states are needed to get quantitatively correct results.

\begin{figure}[h!]
\centering
\mbox{
      \subfigure[Driven 2-cell wire]{\scalebox{0.4}{\includegraphics{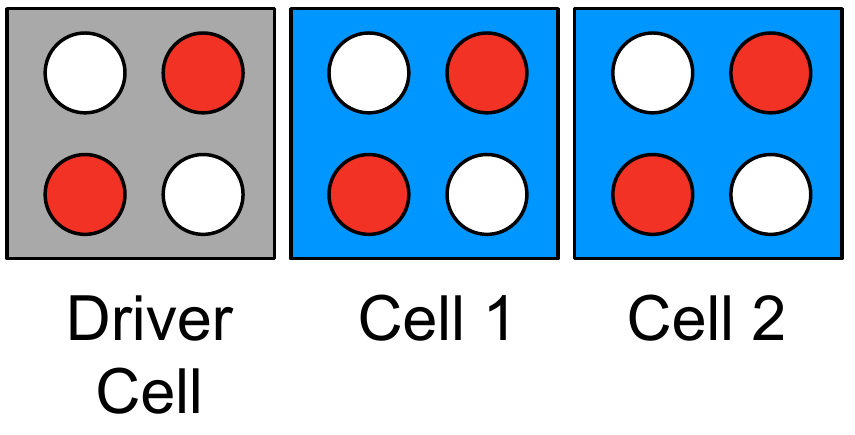}}}
      } \\
      \mbox{
      \subfigure[ICHA]{\scalebox{0.33}{\includegraphics{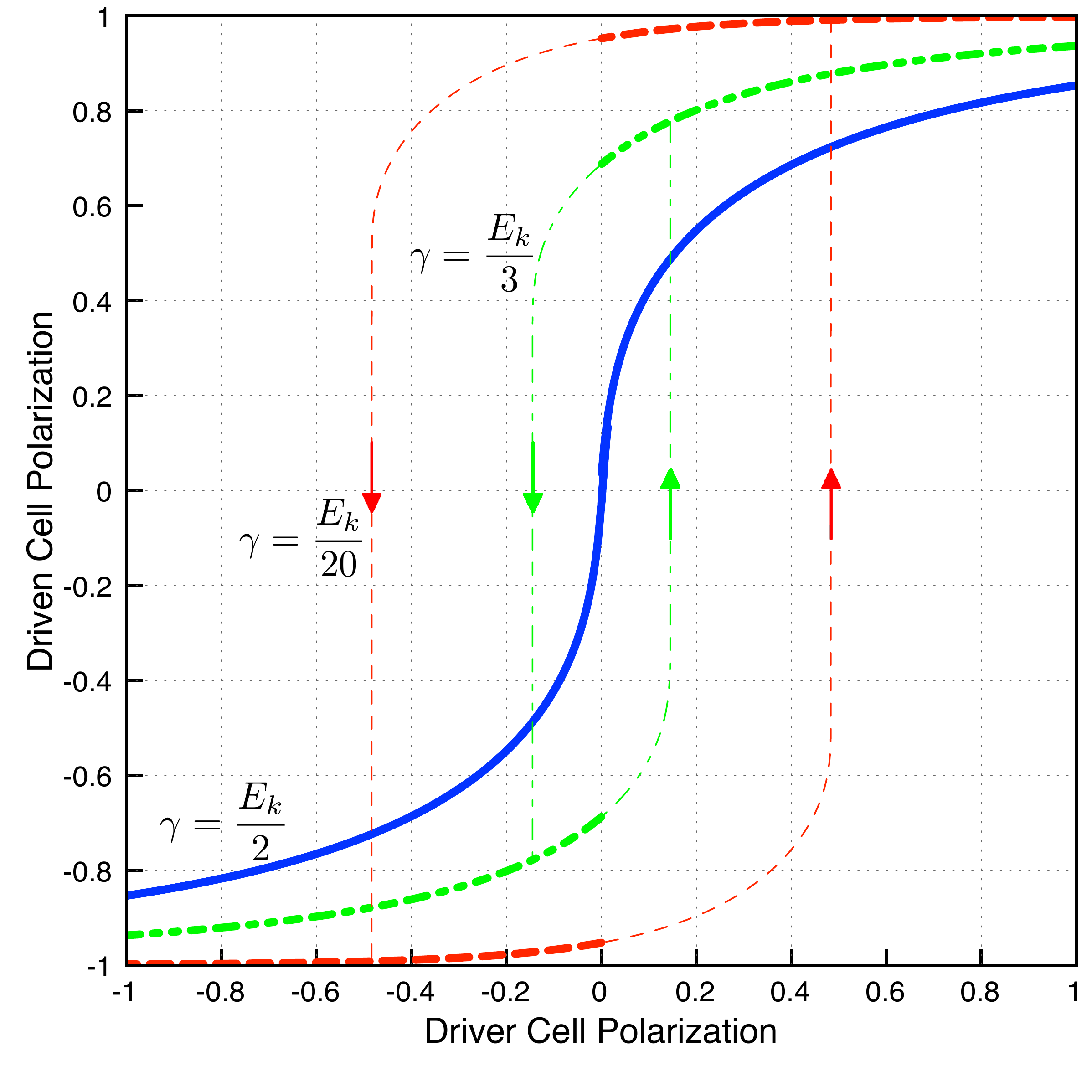}}}
      } 
      \mbox{
      \subfigure[Full QM Model]{\scalebox{0.33}{\includegraphics{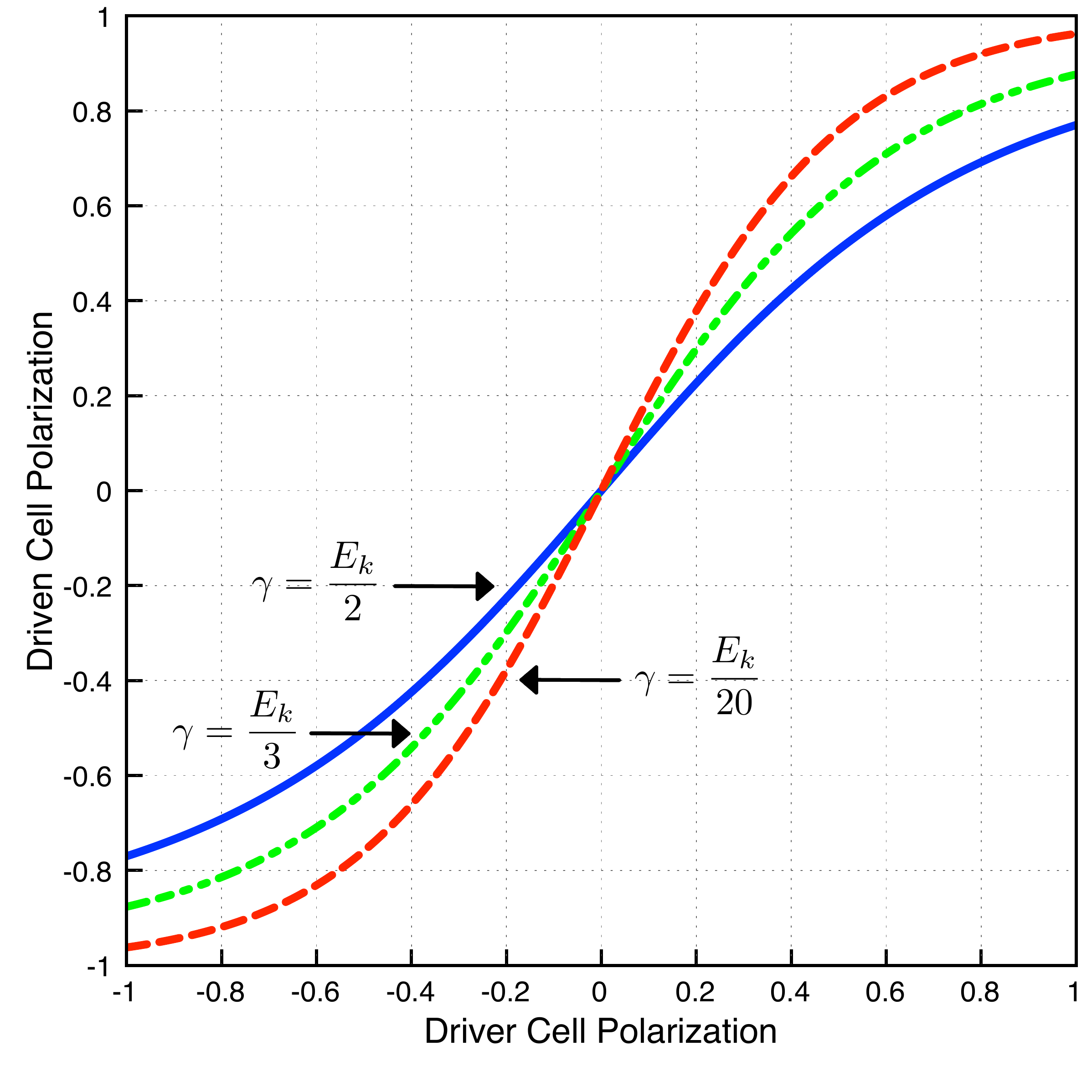}}}
      } 
\caption{Parametric simulations of a driven two-cell wire for different tunnelling rates. (a) The two-cell wire being simulated. (b) Polarization of Cell 1 as a function of the driver cell polarization, calculated using the ICHA.  The ICHA predicts a hysteretic response indicating a memory effect. Where two solutions exist, the one with lowest energy is shown in bold. (c) Polarization of Cell 1 as a function of the driver cell polarization, calculated using the more complete quantum mechanical treatment described by equation \ref{eqn:ising}. This more complete simulation shows no hysteresis. We have taken $k_B T \approx E_k / 4$, which corresponds to $E_k \approx 100~\mathrm{meV}$ for room temperature.}
\label{fig:hysteresis}
\end{figure}

To illustrate the effects of the ICHA on the calculated ground state of QCA arrays, consider the driven two-cell wire shown in Figure~\ref{fig:hysteresis}(a). Two simulations were conducted on the wire; the first using the ICHA, and a second using the more complete quantum mechanical treatment discussed in Section~\ref{sec:two_state}. For each simulation, the driver cell polarization was swept between, $-1 \le P \le +1$, and repeated for cycles with different values of $E_k / \gamma$. The polarization of the first cell in each of these simulations is plotted in Figures~\ref{fig:hysteresis}(b)-(c).

Let us first consider the results from the ICHA simulation. When using the ICHA, one has freedom in choosing an initial guess for the polarizations. The most common method in dynamic simulations is to use as an initial guess at each time step the solution from the previous time step. With this method, Cell 1's response to the driver cell follows the hysteresis curve shown in Figure~\ref{fig:hysteresis}(b). There is a retained polarization even as the driver polarization goes to zero, which is indicative of a memory effect. This type of hysteresis curve is commonly seen in ferromagnets. When a magnetic field is applied to a ferromagnet, electron spins in the ferromagnetic material align themselves with it. For temperatures below the Curie temperature, this alignment is retained when the field is removed, and the material remains (at least partially) magnetized until a magnetic field in the opposite direction is applied. The ICHA predicts an analogous behaviour for QCA cells, with the role of the external magnetic field, in this case, played by the driver cell. When the driver cell is turned on, both the driven cells align themselves with the driver polarization. As the driver cell polarization is removed, the coupling between the two driven cells (through expectation values) allows them to retain part of their polarization even as the driver cell polarization reaches zero. Only after a sufficiently strong driver polarization in the \textit{opposite} direction do both cells switch polarizations. The strength of the driver polarization required to switch the driven cells depends solely on the ratio $\gamma / E_k$; the lower the tunnelling rate, the larger the residual polarization of the driven cells as the driver cell's polarization is removed. 

The other method that can be used with the ICHA is to sample the space of polarizations in search of the self-consistent solution with the lowest energy. For certain driver cell polarizations, there are two self-consistent solutions. In Figure~\ref{fig:hysteresis}(b), the lowest energy solution is shown in bold. If the lowest energy solution is always used, the ICHA predicts a discontinuity in the response of the driven cells, at zero driver cell polarization. As the driver's polarization crosses zero, the cells respond by abruptly ``snapping'' to the other polarization state. The ICHA is not typically used in this way, however.

Simulations conducted using a more complete quantum mechanical treatment show no such hysteresis, as shown in Figure~\ref{fig:hysteresis}(c). Here, as the driver polarization is removed, Cell 1 also relaxes to zero polarization. Thus, there is a ``depolarizing" effect that is not predicted when treating QCA systems using the ICHA. Also, the response is continuous, but still non-linear and with a slope greater than unity at the origin.

The ICHA represents, in a sense, the minimum inclusion of quantum mechanical effects and it will be shown that this assumption can lead to errors when applied to clocked QCA. It is possible to make a less drastic simplification of the system dynamics by including some, but not all, intercell correlations, for example by including only nearest-neighbour pair correlations\cite{Toth:2001-2, Faizal:2008}. Including some of the correlations has been shown to give the correct answer in some cases where the ICHA fails. In Section \ref{sec:depol} we will show a case in which the ICHA fails, and the inclusion of nearest neighbour correlations does \textit{not} help.

\subsection{Relaxation Time Approximation}
A final approximation to be discussed here is the relaxation time approximation, which is commonly used in simulations of QCA dynamics. In the absence of energy dissipation and other decohering effects, a QCA array will evolve coherently according to the Liouville equation for the density matrix, $\hat{\rho}$:
\begin{equation}
\frac{d}{dt}\hat{\rho}(t) = \frac{1}{i\hbar}\left[ \hat{H}(t) , \hat{\rho}(t) \right],
\label{eqn:liou}
\end{equation}
which, for a pure state, is exactly equivalent to the Schr\"{o}dinger equation. The Hamiltonian in Eq. \ref{eqn:liou} may be time-dependent, e.g., when driver cells are switched or when cell parameters are changed to implement clocking. 

Over fairly short time scales, quantum mechanical systems often fall to a thermal steady state~\cite{Timler:2002}. If the QCA system is weakly-coupled to the environment, and the energy transfer between the system and environment is well-described by a Markov process, then at low temperatures, the simplest way to incorporate energy dissipation into a model of QCA dynamics is via the relaxation time approximation~\cite{Toth:2001-2, Timler:2002, Weiss:2008, Bhanja:2009-2}. This is done by adding a dissipation term to Eq.~\ref{eqn:liou}:
\begin{equation}
\frac{d}{dt}\hat{\rho}(t) = \frac{1}{i\hbar}\left[ \hat{H}(t) , \hat{\rho}(t) \right]-\frac{1}{\tau}\left( \hat{\rho}(t) - \hat{\rho}_{ss} \right),
\label{eqn:relax}
\end{equation}
where $\tau$ is a phenomenological time constant, and $\rho_{ss}$ is the steady-state matrix defined as
\begin{equation}
\hat{\rho}_{ss} \equiv \frac{e^{-\hat{H}(t)/k_BT}}{\textrm{Tr}\left\{e^{-\hat{H}(t)/k_BT}\right\}}.
\label{eqn:rhoss}
\end{equation}
Determining the steady state density matrix exactly is tantamount to solving the complete Schr\"{o}dinger equation for the system, and thus comes up against all the difficulties mentioned above. It is therefore usually calculated using the ICHA and the two-state approximation. 

Eq.~\ref{eqn:relax} imposes an exponential approach of the density matrix towards its steady-state value, with a time constant $\tau$. This is one example of a quantum master equation, or an equation of motion for the density matrix, which in this case is phenomenological. It would be useful to derive a quantum master equation based on the microscopic details of the system, its environment, and their interaction. This would show whether or not the form of Eq.~\ref{eqn:relax} is correct, and how the value of $\tau$ relates to microscopic parameters. However, a precise quantum master equation will in general be implementation dependent. In this paper, we treat an idealized QCA that is not tied to a specific implementation. Furthermore, regardless of the specifics of the quantum master equation, the density matrix in Eq.~\ref{eqn:rhoss} will very often represent the real steady state of atomic and molecular systems. We therefore attempt to calculate the correct steady state behaviour, acknowledging that the dynamics and the specific value of $\tau$ will be implementation-dependent.\\

\section{Full Quantum Mechanical Calculations} \label{sec:qm}
Having developed an appreciation for the effects of the ICHA and of the relaxation time approximation on the calculated dynamics of QCA arrays, we will briefly consider the qualitative features of some tractable systems of interacting QCA cells by once again considering the two-state Hamiltonian described in equation~\ref{eqn:ising}. 

The difficulty in solving Eq.~\ref{eqn:ising} can be reduced by assuming only nearest-neighbour coupling so that $E_k^{i,j} = E_k \delta_{i\pm1,j}$ and $E_k^{i,D}=0$ for cells that are not adjacent to the driver labelled $D$. This does not imply that \textit{correlations} beyond nearest neighbours are ignored, however. We also restrict our attention to linear chains of cells.

\begin{figure}[h!]
\centering
\mbox{
      \subfigure[Line Spectra]{\scalebox{0.8}{\includegraphics[width=0.75\textwidth]{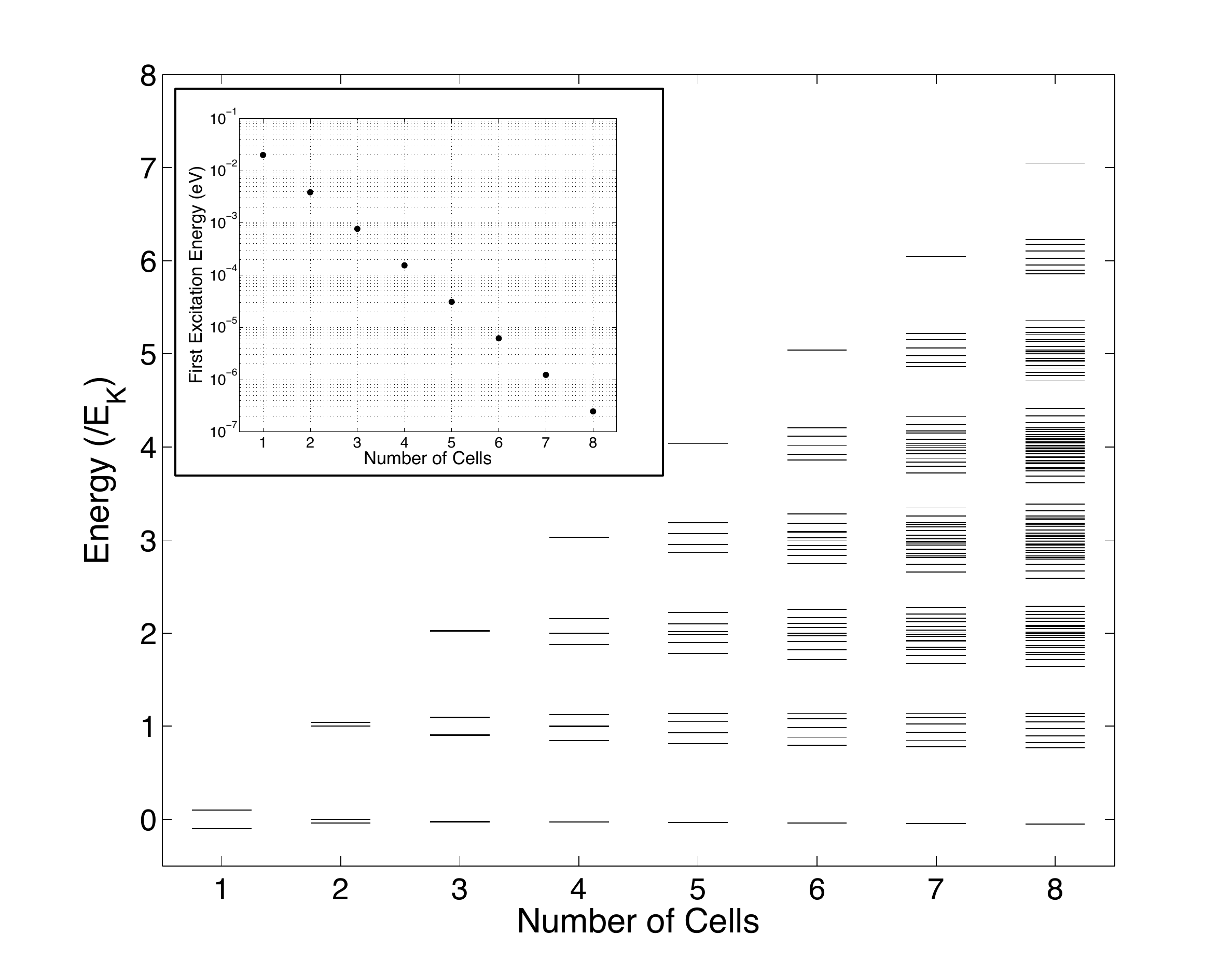}}}
      } \qquad \qquad \\
      \mbox{
      \subfigure[Analogous Double Well Systems]{\scalebox{0.4}{\includegraphics{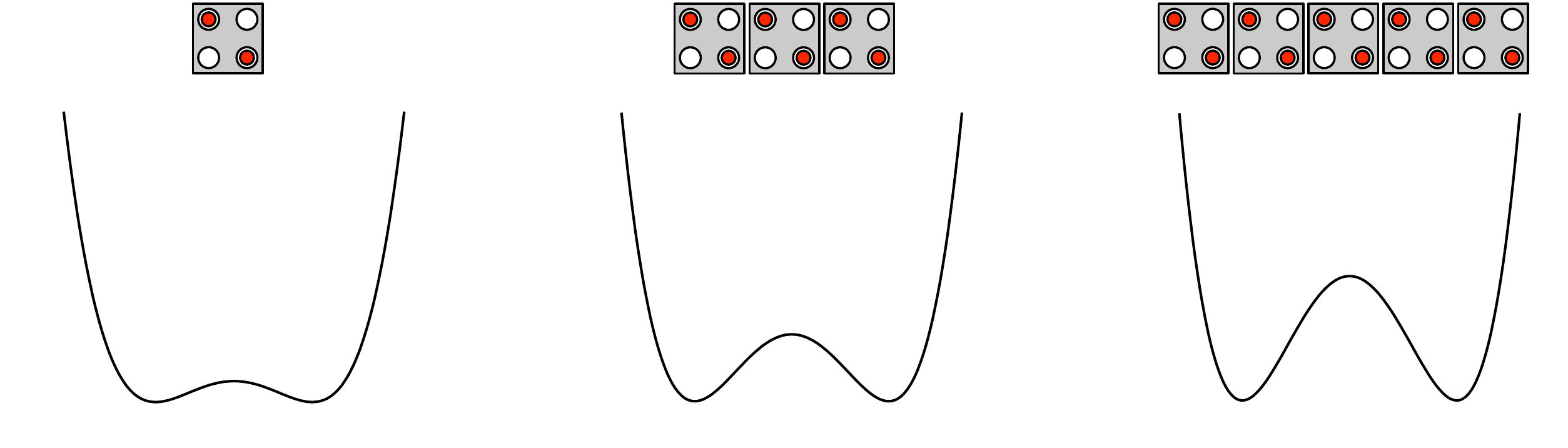}}}
      } 
\caption{(a) Spectra of unbiased lines of interacting QCA cells, ranging in length from one to eight cells, with $\gamma / E_k = 0.1$. The spectra are solutions of Eq.~\ref{eqn:ising}, with a constant $(N-1)E_k/2$ added for ease of interpretation. The inset shows the difference in energy between the two lowest energy levels for each line. (b) Conceptually, the array can be viewed as a single two state system with a barrier that increases with the number of cells.}
\label{fig:linespec}
\end{figure}

The spectrum for an unbiased line of $N$ (ranging from 1 to 8) QCA cells, within the two-state approximation is shown in Figure~\ref{fig:linespec}(a). For a line of $N$ cells, there are $2^N$ eigenstates. Under the conditions required for QCA-based computing, that is $\gamma << E_k$, the energy levels come in clusters separated roughly by $E_k$. There are always two non-degenerate lowest-energy states. In the case of a single cell, these are in fact the only two states, and their separation is exactly $2\gamma$. They correspond to the symmetric ($|\psi_s\rangle = \frac{1}{\sqrt{2}}(|0\rangle + |1\rangle)$) and anti-symmetric ($|\psi_a\rangle = \frac{1}{\sqrt{2}}(|0\rangle - |1\rangle)$) combinations of the polarization states (with the former being the ground state). We note that the polarization basis vectors, $|0\rangle$ and $|1\rangle$, are \textit{not} energy eigenstates of the undriven cell. If a single quantum measurement of $\hat{\sigma}_z$ is carried out on a QCA cell in its ground state (or in its first excited state for that matter), the outcome will yield either $-1$ or $+1$, with equal probability, \textit{i.e.}, $P=0$. In this sense, one can say that the ground state of a single unbiased cell carries no information. Only in the limit where $\gamma \rightarrow 0$ do the polarization basis vectors become valid energy eigenstates.

For longer lines, the separation between the two lowest-lying energy eigenstates becomes smaller and they represent entangled states. Specifically, they are the symmetric and anti-symmetric combinations of the state with all cell polarizations aligned along one diagonal, and the state with them aligned along the other. That is, $|\psi_{s,a}\rangle \approx \frac{1}{\sqrt{2}}(|000\dots0\rangle \pm |111\dots1\rangle)$. The equality becomes exact in the limit where $E_k / \gamma \rightarrow \infty$. 

Again, we note that the ``aligned'' states, $|000\dots0\rangle$ and $|111\dots1\rangle$, are not the energy eigenstates. The energy eigenstates are in fact superpositions of the aligned states, and the polarization of any cell in a line in its ground state (or first excited state) is $P=0$. This indicates that the eigenstates do not carry information. The inset of Figure \ref{fig:linespec}(a) shows that the splitting between these two lowest-lying states decreases exponentially. Specifically, each added cell causes the splitting to decrease by a factor of $\sim E_k/\gamma$. Only in the limit where this splitting goes to zero (\textit{i.e.} an infinite number of interacting cells), do the aligned states become energy eigenstates\cite{Pfeuty:1969}. With the decrease in the splitting of the two lowest energy levels, comes an increase in the time required for coherent tunnelling from one polarization state to the other. For low temperatures, we can think of the group of cells as a two-state system, analogous to a double well, as depicted in Figure \ref{fig:linespec}(b). As the length of the line increases, the barrier separating the two aligned states increases. It follows that longer lines acquire increased bistability. However, at finite temperature, thermal fluctuations may cause excitations and an eventual approach to the unpolarized steady state. 

\begin{figure}[h!]
\centering
\includegraphics[width=0.75\textwidth]{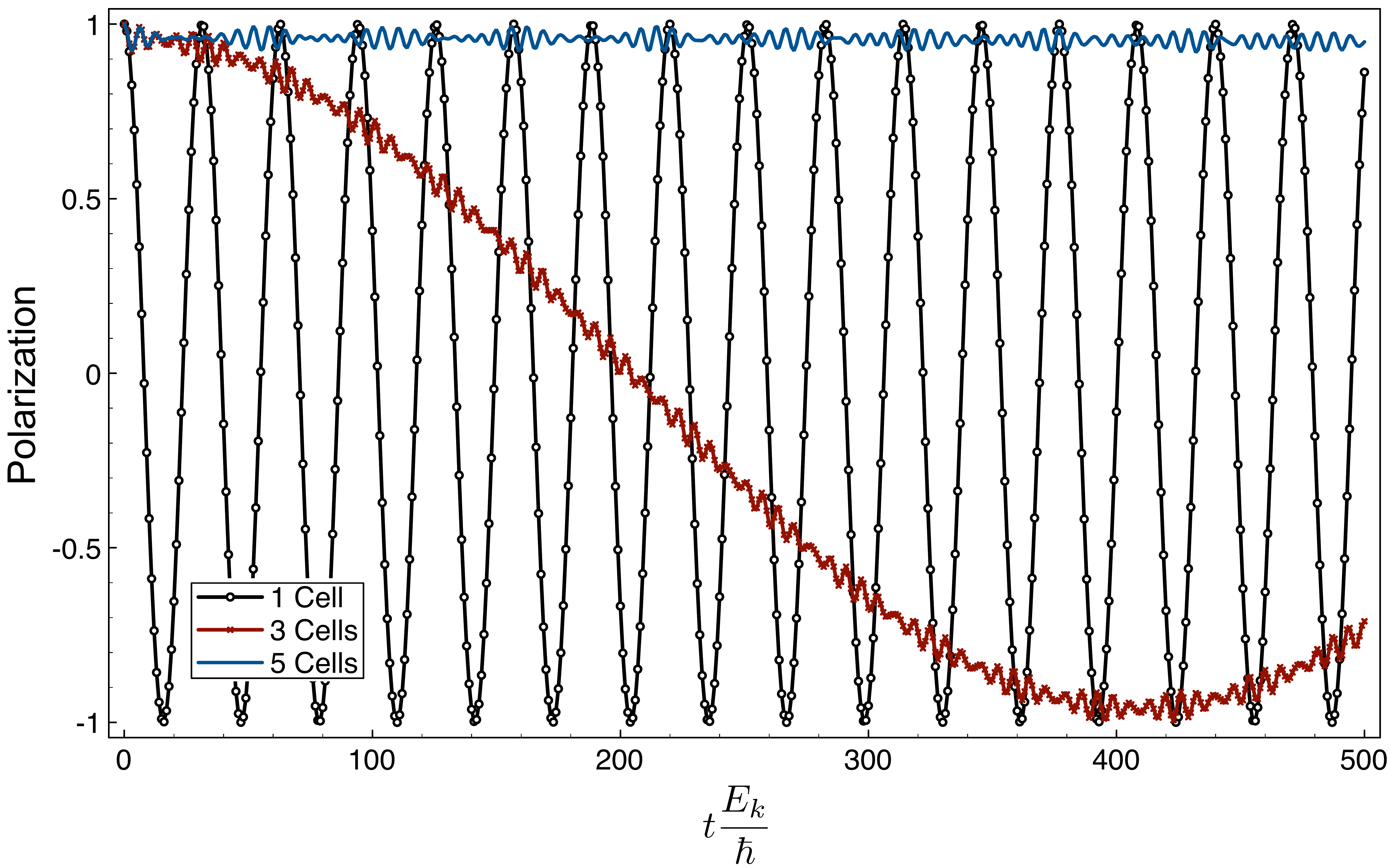}
\caption{Coherent oscillations of the first cell in unperturbed one, three, and five cell lines. The cells are all initially in the $P=1$ state. In each line, the cells oscillate together, so the polarizations of the first cell gives a good representation of the polarization of all its neighbours. The fast oscillations in the 3 and 5 cell lines are due to a small component of higher-energy states, which manifest as kinks propagating and reflecting through the line. $\gamma=10~\textrm{meV}$ and $E_k = 100~\textrm{meV}$.}
\label{fig:cohosc}
\end{figure}

Figure \ref{fig:cohosc} shows the unitary time evolution of unperturbed lines of one, three, and five cells initially in an aligned state in each case. All the cells in the wire oscillate in unison between the two polarization states. This simulation represents the limit of infinite relaxation time (\textit{i.e.}, no energy dissipation or loss of phase coherence). Since the longer lines exhibit increased bistability, the polarization state can be maintained for an arbitrary length of time by increasing the number of cells in the line. For single cells and small groups at the atomic scale, coherent oscillations will likely be much faster than the measurement time, and therefore will lead to a loss of classical information. For larger groups, the loss of information will likely be limited by environment-induced phase decoherence and energy relaxation. 

Finally, we emphasize that short unbiased lines, like individual cells, have a \textit{unique} ground state. This contrasts with the commonly made assertion that cells and lines have two degenerate states\cite{Blair:2011}, namely, the aligned states. The ground state is an entangled state, and is therefore not accessible to the ICHA. Based on this observation, as well as the above-mentioned shortcomings of the ICHA in predicting the behaviour of even very short lines, we are led to a more in-depth investigation of QCA line dynamics, particularly with regard to clock zones which are spatially separated from any driver cells.

\section{Loss of Polarization in Isolated Bit Packets} \label{sec:depol}
As discussed in Section~\ref{sec:ICHA}, the ICHA predicts a latching mechanism within a line of interacting cells that allows them to polarize (and retain this polarization) in the absence of a fixed driver cell. This phenomenon is found to be an artifact of the ICHA, not necessarily representative of actual dynamics. Consider a single, unbiased QCA cell. At low temperatures, the QCA cell will relax to its ground state, $|\psi_s\rangle = \frac{1}{\sqrt{2}}(|0\rangle + |1\rangle)$, and thus will have a polarization, $P = 0$. At higher temperatures, interactions with the environment can cause the QCA cell to be in a mixed state, and it is best described by a density matrix. If the steady state density matrix of the system is the one described in Eq.~\ref{eqn:rhoss}, then it is easy to show that the expectation value of $\sigma_z$ is zero, and therefore the polarization is also zero. For longer lines, the same reasoning applies, the only differences being that $|\psi_{s,a}\rangle$ represent the entangled states described in Section \ref{sec:qm}, and the splitting in their energies is decreased. It can be shown that the polarization of any cell in the unbiased line is zero if the line is in the steady state described by Eq.~\ref{eqn:rhoss}.

\begin{figure}[h!]
\centering
\mbox{
      \subfigure{\scalebox{0.25}{\includegraphics{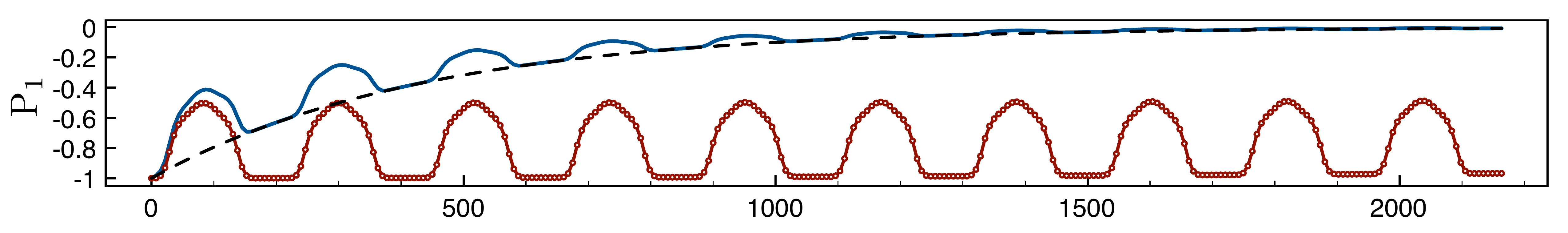}}}
      } \qquad \qquad
     \mbox{
      \subfigure{\scalebox{0.25}{\includegraphics{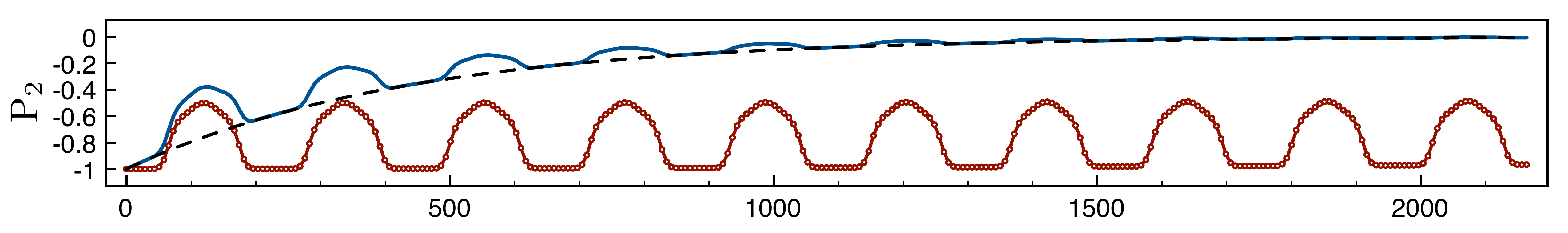}}}
      } \qquad \qquad
      \mbox{
      \subfigure{\scalebox{0.25}{\includegraphics{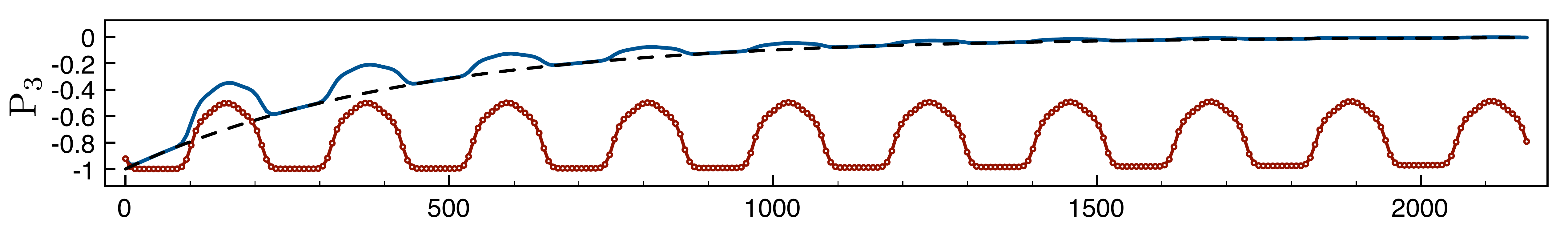}}}
      } \qquad \qquad
      \mbox{
      \subfigure{\scalebox{0.25}{\includegraphics{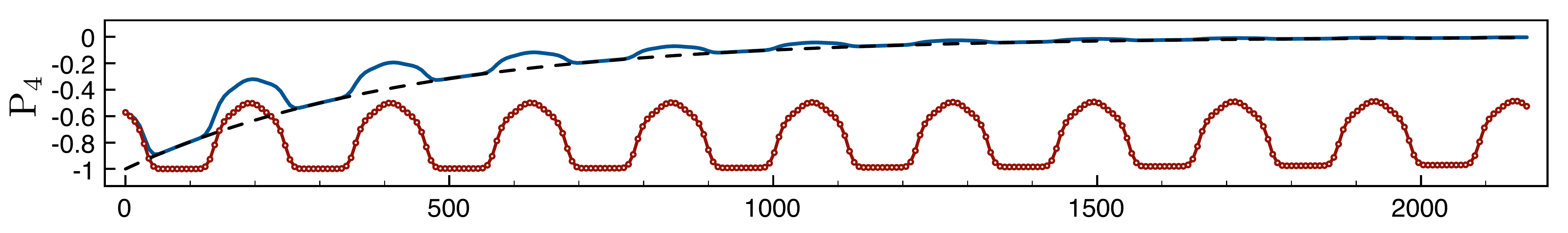}}}
      } \qquad \qquad
      \mbox{
      \subfigure{\scalebox{0.25}{\includegraphics{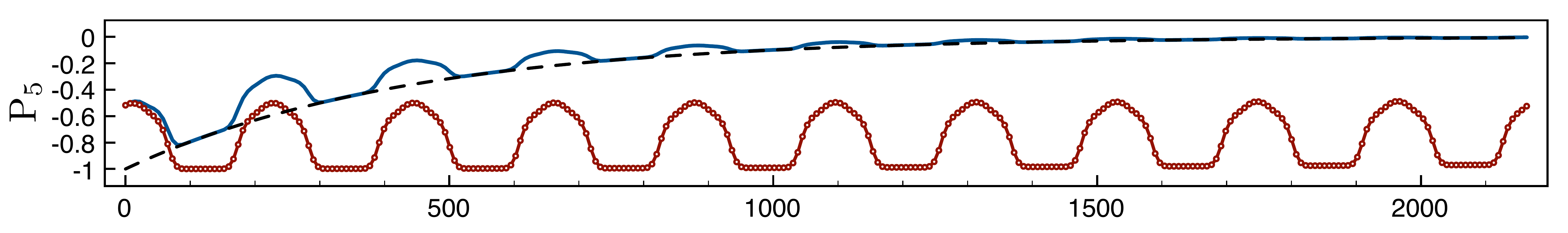}}}
      } \qquad \qquad
      \mbox{
      \subfigure{\scalebox{0.25}{\includegraphics{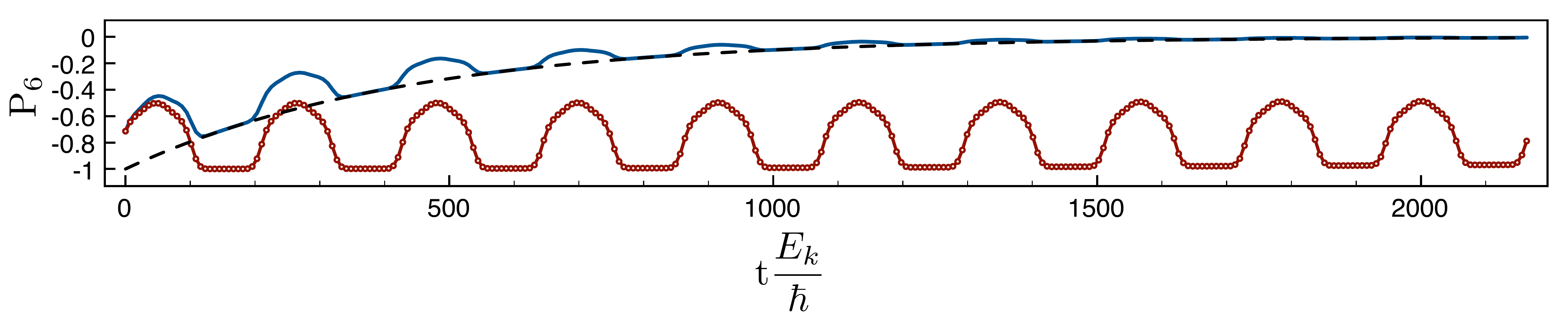}}}
      } 
\caption{Simulations of a six cell line with periodic boundary conditions. The value of $\gamma$ is adiabatically raised and then lowered between 1 < $\gamma$ < 200 meV, with a period of $217\hbar/E_k$, with a different phase for each cell. The circled line shows the polarization of each cell in the absence of any energy dissipation. The solid line shows the polarization of each cell with an energy relaxation time of $434\hbar/E_k$. For reference, a decaying exponential as a function of $\tau$ is shown with the dashed line.}
\label{fig:ring}
\end{figure}

\begin{figure}[h!]
\centering
\mbox{
      \subfigure{\scalebox{0.25}{\includegraphics{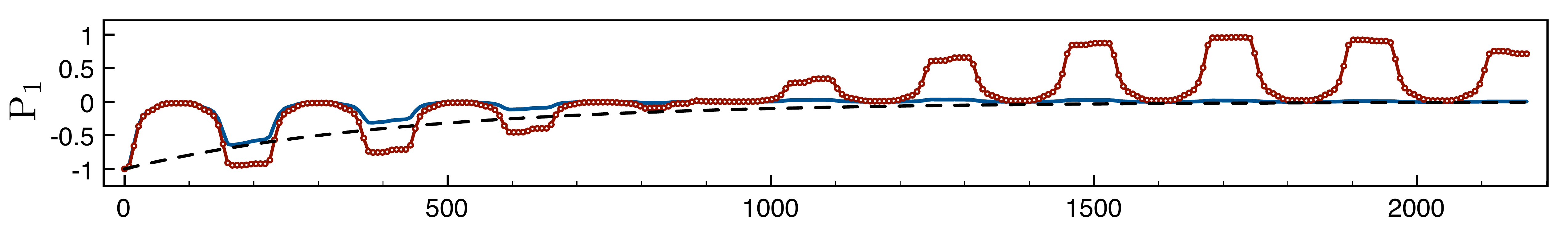}}}
      } \qquad \qquad
     \mbox{
      \subfigure{\scalebox{0.25}{\includegraphics{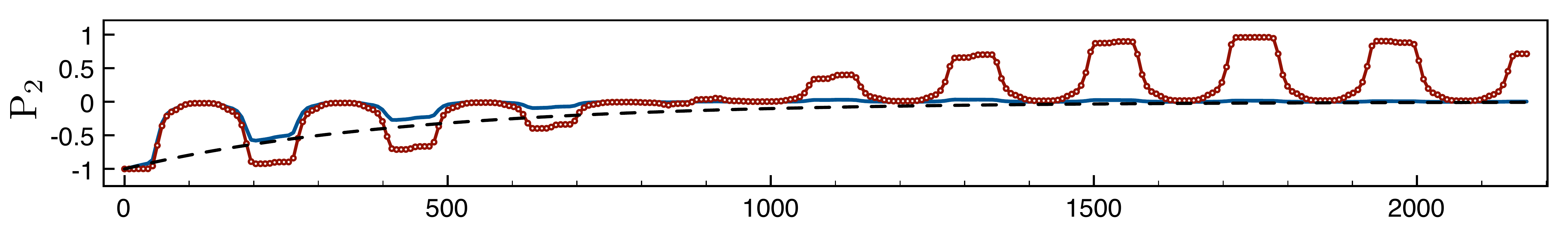}}}
      } \qquad \qquad
      \mbox{
      \subfigure{\scalebox{0.25}{\includegraphics{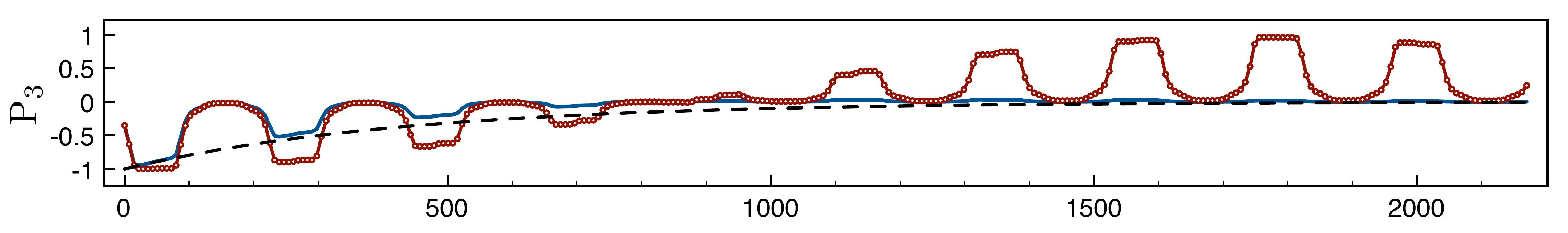}}}
      } \qquad \qquad
      \mbox{
      \subfigure{\scalebox{0.25}{\includegraphics{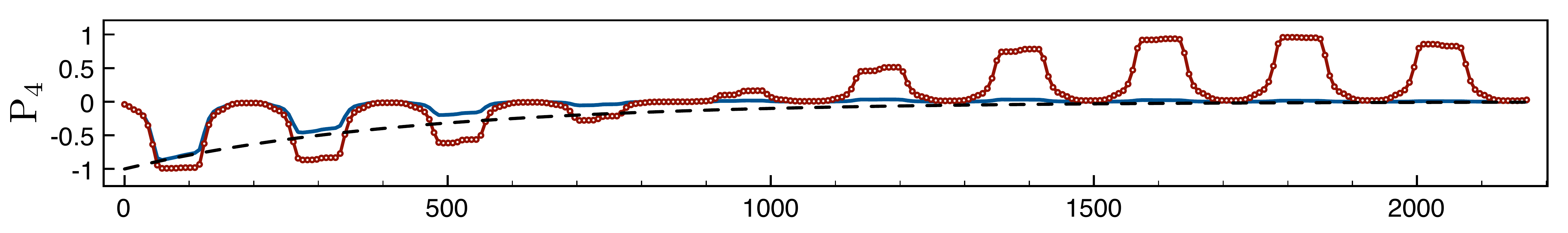}}}
      } \qquad \qquad
      \mbox{
      \subfigure{\scalebox{0.25}{\includegraphics{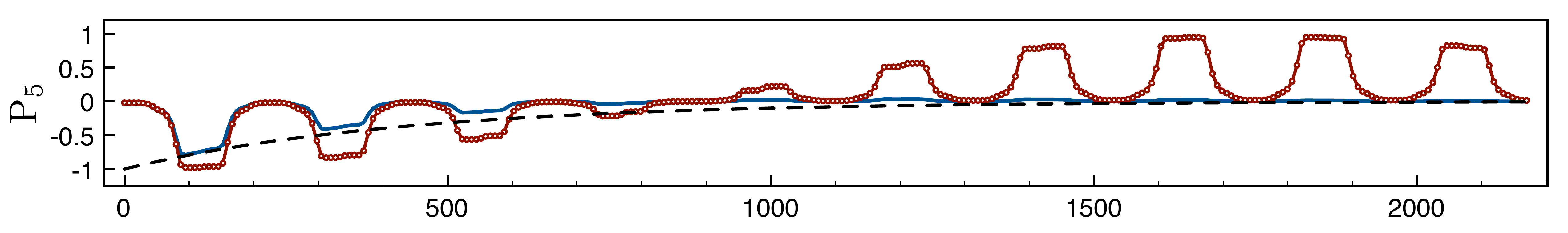}}}
      } \qquad \qquad
      \mbox{
      \subfigure{\scalebox{0.25}{\includegraphics{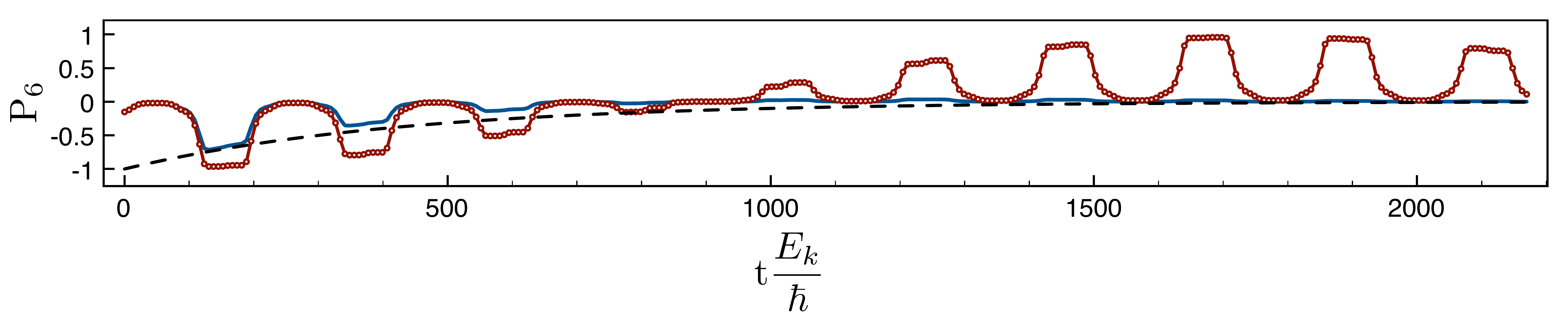}}}
      } 
\caption{Simulations of a six cell line with periodic boundary conditions. The value of $\gamma$ is adiabatically raised and then lowered between 1 < $\gamma$ < 1000 meV, with a period of $217\hbar/E_k$, with a different phase for each cell. The circled line shows the polarization of each cell in the absence of any energy dissipation. The solid line shows the polarization of each cell with an energy relaxation time of $434\hbar/E_k$. For reference, a decaying exponential as a function of $\tau$ is shown with the dashed line. Half a period of a coherent oscillation can be seen over the course of the simulation, flipping the polarization state from negative to positive.}
\label{fig:ring2}
\end{figure}

Figures~\ref{fig:ring} and~\ref{fig:ring2} show the results of two simulations of six cells in a one-dimensional chain, with periodic boundary conditions and nearest neighbour coupling only. $\gamma$ is modulated between $1$ meV and $200$ meV for the first simulation, and between $1$ meV and $1000$ meV in the second. $E_k=108.5$ meV in both simulations. The Hamiltonian in equation \ref{eqn:ising} is used, so that intercellular correlations are completely included. In both simulations, the tunneling barriers in cells 1, 2, and 3 are initially high (meaning that $\gamma$ is low) while the tunneling barriers in cells 4, 5, and 6 are low (so that $\gamma$ is high). As time progresses, the tunneling barriers in cell 4 are raised, allowing it to polarize, and the barriers in cell 1 are lowered, which causes it to depolarize. Next, cell 5 has its barriers raised as the barriers in cell 2 are lowered, and so on. Because of the periodic boundary conditions, the bit packet moves cyclically through the six cells. Throughout the process, the Hamiltonian changes quasi-adiabatically, by which we mean that $d\gamma/dt \ll E_k^2 / \hbar$. For our initial state, we create a nearly full negative polarization in the three active cells by taking the normalized sum of the two lowest energy eigenstates. The resulting cell polarizations are plotted as a function of time with and without dissipation in each case. A third curve plotting the exponential decay, proportional to $e^{-t/\tau}$, is also shown in each plot as a reference. 

In the simulation of Figure~\ref{fig:ring}, as time progresses, the bit packet travels along the periodic line, and a polarization is maintained when there is no dissipation. Because the Hamiltonian is changed quasi-adiabatically, no kinks are created as the bit packet moves, and the bit packet evolves qualitatively the same as it would if it were stationary. Because the tunneling barriers are never lowered completely ($\gamma$ has a maximum value of $200$ meV), the cells never reach zero polarization. The interaction between the three active cells at a given time, plus the residual polarization in the ``inactive" cells, slows the coherent oscillations to the point where the bit packet is fully polarized over the period of the simulation, as long as there is no dissipation. However, the steady state still has zero polarization in all cells, so the bit packet loses polarization exponentially when relaxation is included in the model. The slow oscillation frequency demonstrated in this simulation simply permits the cells to maintain their polarizations over several clock cycles. 

In the simulation of Figure~\ref{fig:ring2}, since the tunnelling barriers are lowered enough to completely depolarize the cells ($\gamma$ is now allowed as high as $1000$ meV), fewer cells are ``on" at any given time, and the coherent oscillations from negative to positive polarization state are sufficiently fast to be observed over the time scale of the simulation. As in Figure~\ref{fig:ring}, the dissipation of energy does indeed bring the system exponentially to its steady state of zero polarization. In this simulation, because of the visible coherent oscillations, the polarization does not follow the decaying exponential as it does in Figure~\ref{fig:ring}, but instead the amplitude of the oscillation is proportional to the exponential decay.

\section{Discussion} \label{sec:disc}
The simulations in Section~\ref{sec:depol} show that the ICHA does not always provide a good approximation to the full quantum mechanical model of an isolated array of QCA cells. In the case of bit packets which are not coupled to a driving cell, the exponential relaxation causes any initial polarization to be lost. Within the framework of equations~\ref{eqn:ising},~\ref{eqn:relax}, and~\ref{eqn:rhoss}, the maximum time for a clocked computation will be limited by the loss of classical information either through energy relaxation or through oscillations of the type shown in Figures 5 and 7. In principle, coherent oscillations do not imply a loss of information (the information can be retrieved by a carefully timed measurement), however, the oscillations may be much more rapid than the duration of a measurement, \textit{e.g.} in the case of a small number of cells, and will in general suffer phase decoherence, which will erase even this phase information. All the simulations presented here were done within the two-state approximation of equation \ref{eqn:ising}, which is an approximation of equation \ref{eqn:hubbard}. As mentioned previously, working directly with equation \ref{eqn:hubbard} is very computationally expensive. However, based on the conceptual understanding outlined in this paper, in addition to other computational results on very small groups of cells (not shown here), we believe that these new features of clocked QCA are not an artifact of the two-state approximation, but are the result of the inclusion of intercellular correlations and the use of the relaxation time approximation. 

Previous simulations have used the ICHA to predict the latching of lines and the successful propagation of clocked pulses across QCA arrays (even when tunnelling rates were varied by only one order of magnitude~\cite{Timler:2002}). These have seen some degree of validation from experimental work on lithographically defined QCA-like systems\cite{Orlov:2001}\cite{Orlov:2003}. The results presented here show that these experiments do not embody the dynamics of equations \ref{eqn:ising} and \ref{eqn:relax}. The use of intermediate quantum dots and multiple tunnel junctions in these experiments makes it possible to tune the tunneling time from $10ps$ to $\sim 3000s$\cite{Orlov:2001}. This drastic change in tunneling rates amounts to a crossover from a regime where quantum tunneling is important to one where electrons are completely localized. The normal relaxation dynamics are suppressed because the system is strongly coupled to the environment through $\sigma_z$, which has the effect of localizing charge\cite{Leggett:1987}. In molecular and atomic implementations, it is likely that the range of tunneling rates that allows this crossover from quantum dynamics to classical dynamics by directly modulating tunnel barriers, will not be achievable. A more promising approach to achieving such a crossover would be to change the internal dynamics of a bit-packet by changing its size. In section \ref{sec:qm}, we showed that longer lines exhibit slower dynamics and greater bistability. This suggests that for sufficiently large bit packets, there may be sufficient bistability that the coupling to the environment results in localization of charge, even with a limited range of tunneling rates available. The precise tunneling rate and bit packet size that will allow localization of charge with sufficiently slow relaxation dynamics will be determined by the nature and strength of the coupling of the environment to the QCA system, which will in turn compete with the QCA system's internal dynamics. A detailed analysis will need to take into account the microscopic details of a specific implementation.

The simulations performed in this work clearly identify depolarization due to quantum correlations as a critical issue for classical computation using clocked QCA at the molecular and atomic scale. To our knowledge, this  phenomenon has not been acknowledged in any of the previous work on QCA. However, it is important to note that this loss of polarization occurs only when a set of cells becomes isolated from a perturbing influence such as a fixed driver cell, and thus, QCA systems operating within a single clocking zone will not experience such an effect. 

\section{Conclusion} \label{sec:conc}
In this paper, we have assessed some limitations of conventional approaches to QCA simulation. Full quantum mechanical calculations show that the ground state of an unbiased cell, or of a line of cells, is a superposition of the two fully aligned states, and thus holds no polarization and carries no information. When the assumption of exponential relaxation to a thermal steady state is made, we find that cells, or groups of interacting cells, lose their polarization over time unless they are influenced by a fixed driver. This is the case even if the cells start with an initial polarization before being decoupled from a fixed driver cell. This depolarization effect was not predicted in previous QCA simulations, which had often predicted a false latching mechanism among cells that would allow them to retain their polarization even in the absence of a driver cell. We have found that this failure is related to the ICHA which neglects correlations and shows hysteresis in array polarization. Although lithographic QCA systems have managed to avoid this problem thanks to their inherently long tunnelling times, the molecular and atomic implementations of QCA required for room temperature operation will likely behave in a more purely quantum mechanical way, so that the solutions of the many-cell Hamiltonian need to be included. Only an appropriate and sufficiently strong interaction of a QCA array with its environment will make clocked QCA possible.

While these findings do not affect the original concept of ground state computing with QCA, they may require a reconsideration of QCA architecture, specifically relating to clocking and memory devices. Because the simulations presented here are still for a very highly idealized model of QCA behaviour, and ignore, among other things, the specifics of the interaction with the environment, we do not claim that clocking or memory are impossible in QCA. Effective clocking requires a tuneable change from quantum mechanical behaviour to classical. It remains to be shown how and if this can be achieved at the molecular scale. This underscores the need for a more sophisticated theory of QCA operation, which should include implementation-specific dynamics beyond the phenomenological relaxation time approximation. 

\bibliographystyle{ieeetr}
{
\bibliography{QCAcorr}
}

\end{document}